\pdfoutput=1

\documentclass[letterpaper]{article} % DO NOT CHANGE THIS
\usepackage{aaai23}  % DO NOT CHANGE THIS
\usepackage{times}  % DO NOT CHANGE THIS
\usepackage{helvet}  % DO NOT CHANGE THIS
\usepackage{courier}  % DO NOT CHANGE THIS
\usepackage[hyphens]{url}  % DO NOT CHANGE THIS
\usepackage{graphicx} % DO NOT CHANGE THIS
\usepackage{natbib}  % DO NOT CHANGE THIS AND DO NOT ADD ANY OPTIONS TO IT
\usepackage{caption} % DO NOT CHANGE THIS AND DO NOT ADD ANY OPTIONS TO IT
\usepackage{algorithm}
\usepackage{amsmath,amsthm,amsfonts,amssymb,bm}

\usepackage{mathrsfs}
\usepackage{multirow}
\usepackage[table,xcdraw]{xcolor}
\usepackage{booktabs}
\usepackage{graphicx} 
\usepackage{xspace} 
\usepackage{algpseudocode}
\usepackage{algorithmicx}
\usepackage[T1]{fontenc} % Ensure that the English font is bold and effective
\usepackage{bm}
\usepackage{enumitem}
\usepackage{varwidth}

\usepackage{newfloat}
\usepackage{listings}
\DeclareCaptionStyle{ruled}{labelfont=normalfont,labelsep=colon,strut=off} % DO NOT CHANGE THIS
\floatstyle{ruled}
\newfloat{listing}{tb}{lst}{}
\floatname{listing}{Listing}

\usepackage{subfig}

\graphicspath{{figs/}}

\newcommand{\ours}{PC-Attack\xspace}
\newcommand{\oursstar}{PC-Attack$^*$\xspace}

\title{Practical Cross-System Shilling Attacks with Limited Access to Data}

\author {
    Meifang Zeng\textsuperscript{\rm 1},
    Ke Li\textsuperscript{\rm 2},
    Bingchuan Jiang\textsuperscript{\rm 2},
    Liujuan Cao\textsuperscript{\rm 1},
    Hui Li\textsuperscript{\rm 1}\thanks{Hui Li is the corresponding author.}\\
}
\affiliations {
    \textsuperscript{\rm 1} School of Informatics, Xiamen University \\
    \textsuperscript{\rm 2} PLA Strategic Support Force Information Engineering University \\
    zengmeifang@stu.xmu.edu.cn, \{like19771223,\,jbc021\}@163.com, \{caoliujuan,\,hui\}@xmu.edu.cn
}

\begin{document}

\maketitle

\begin{abstract}
In shilling attacks, an adversarial party injects a few fake user profiles into a Recommender System (RS) so that the target item can be promoted or demoted. Although much effort has been devoted to developing shilling attack methods, we find that existing approaches are still far from practical. In this paper, we analyze the properties a practical shilling attack method should have and propose a new concept of Cross-system Attack. With the idea of Cross-system Attack, we design a Practical Cross-system Shilling Attack (PC-Attack) framework that requires little information about the victim RS model and the target RS data for conducting attacks. PC-Attack is trained to capture graph topology knowledge from public RS data in a self-supervised manner. Then, it is fine-tuned on a small portion of target data that is easy to access to construct fake profiles. Extensive experiments have demonstrated the superiority of PC-Attack over state-of-the-art baselines. Our implementation of PC-Attack is available at https://github.com/KDEGroup/PC-Attack.
\end{abstract}

\section{Introduction}
\label{sec:intro}

Recommender System (RS) has become an essential tool in various online services.
However, its prevalence also attracts attackers who try to manipulate RS to mislead users for gaining illegal profits.
Among various attacks, \emph{Shilling Attack} is the most subsistent and profitable one~\citep{abs-2206-11433}.
RS allows users to interact with the system through various operations such as giving ratings or browsing the page of an item.
In shilling attacks, an adversarial party injects a few fake user profiles into the system to hoax RS so that the target item can be promoted or demoted~\citep{GunesKBP14}.
This way, the attacker can increase the possibility that the target item can be viewed/bought by people or impair the competitors by demoting their products.
In experiments, shilling attacks are able to spoof real-world RS, including Amazon, YouTube and Yelp~\citep{XingMDSFL13,YangGC17}.
In practice, services of various large companies were affected by shilling attacks~\citep{LamR04}.
Studying how to spoof RS has become a hot direction~\cite{DeldjooNM21} as it gives insights into the defense against malicious attacks.

Although much effort has been devoted to developing new shilling attack methods~\cite{GunesKBP14,DeldjooNM21}, 
we find \emph{existing shilling attack approaches are still far from practical}\footnote{The detailed analysis is provided in Sec.~\ref{sec:analysis}.}.
The main reason is that most of them require the complete knowledge of the RS data, which is not available in  real shilling attacks.
A few works study attacking using incomplete data~\citep{ZhangTLSYZG21} or transferring knowledge from other sources to attack the victim RS~\citep{FanDZ0LWT021}.
Nevertheless, they still request a large portion of the target data or assume other data sources share some items with the victim RS.

In this paper, we study the problem of designing a \emph{practical} shilling attack approach.
We believe a practical shilling attack method should have the following nice properties:
\begin{itemize}
\item \textbf{Property 1:} Do not require any prior knowledge of the victim RS (e.g., model architecture or parameters in the model).

\item \textbf{Property 2:} When training the attacker, other data sources (e.g., public RS datasets) instead of the data of the victim RS can be used. Do not assume the training data contain any users or items that exist in the victim RS.

\item \textbf{Property 3:} The attacker should use as little information of the data in the victim RS as possible. Required information should be easy to access in practice.
\end{itemize}

\begin{table*}[t]
\centering
\resizebox{1\linewidth}{!}{
\begin{tabular}{ccccccccc}
\hline
\multirow{2}{*}{Category}     & \multirow{2}{*}{Method}                                         & \multicolumn{3}{c}{Knowledge}                                                                                                                                               & \multirow{2}{*}{\begin{tabular}[c]{@{}c@{}}Do not train with\\ a surrogate RS\end{tabular}} & \multirow{2}{*}{\begin{tabular}[c]{@{}c@{}}Do not require\\ multiple queries\end{tabular}} & \multirow{2}{*}{\begin{tabular}[c]{@{}c@{}}Cross-domain\\ attack\end{tabular}} & \multirow{2}{*}{\begin{tabular}[c]{@{}c@{}}Cross-system\\ attack\end{tabular}} \\ \cline{3-5}
                              &                                                                 & \begin{tabular}[c]{@{}c@{}}Target\\ Data\end{tabular} & \begin{tabular}[c]{@{}c@{}}RS\\ Architecture\end{tabular} & \begin{tabular}[c]{@{}c@{}}RS\\ Parameters\end{tabular} &                                                                                             &                                                                                            &                                                                                &                                                                                \\ \hline
\multirow{2}{*}{Optimization} & PGA and SGLD                                                    & $m\cdot n$                                            & \checkmark                                                & \checkmark                                              & \checkmark                                                                                  & \checkmark                                                                                 & ×                                                                              & ×                                                                              \\
                              & RevAdv and RAPU                                                 & $m\cdot n$                                            & ×                                                         & ×                                                       & ×                                                                                           & \checkmark                                                                                 & ×                                                                              & ×                                                                              \\ \hline
\multirow{3}{*}{GAN}          & TrialAttack                                                     & $m\cdot n$                                            & \checkmark                                                & ×                                                       & ×                                                                                           & \checkmark                                                                                 & ×                                                                              & ×                                                                              \\
                              & Leg-UP                                                          & $m\cdot n$                                            & ×                                                         & ×                                                       & ×                                                                                           & \checkmark                                                                                 & ×                                                                              & ×                                                                              \\
                              & \begin{tabular}[c]{@{}c@{}}DCGAN, AUSH\\ and RecUP\end{tabular} & $m\cdot n$                                            & ×                                                         & ×                                                       & \checkmark                                                                                  & \checkmark                                                                                 & ×                                                                              & ×                                                                              \\ \hline
\multirow{3}{*}{RL}           & PoisonRec                                                       & $e\cdot n \cdot k$                                    & ×                                                         & ×                                                       & \checkmark                                                                                  & ×                                                                                          & ×                                                                              & ×                                                                              \\
                              & LOKI                                                            & $m\cdot n$                                            & ×                                                         & ×                                                       & ×                                                                                           & \checkmark                                                                                 & ×                                                                              & ×                                                                              \\
                              & CopyAttack                                                      & $e\cdot n \cdot k$                                    & ×                                                         & ×                                                       & \checkmark                                                                                  & ×                                                                                          & \checkmark                                                                     & ×                                                                              \\ \hline
KD                            & Model Extraction Attack                                         & $c \cdot k$                                           & \checkmark                                                & ×                                                       & \checkmark                                                                                  & ×                                                                                          & ×                                                                              & ×                                                                              \\ \hline
                              & \ours                                                           & $p \cdot m\cdot n$                                    & ×                                                         & ×                                                       & \checkmark                                                                                  & \checkmark                                                                                 & \checkmark                                                                     & \checkmark                                                                     \\ \hline
\end{tabular}
}
\caption{Comparisons of shilling attack approaches. Reference of each method can be found in Sec.~\ref{sec:related}. $m$ and $n$ indicate the numbers of users and items, respectively. $p$ is the maximum percentage of the target data that \ours requires. $e$, $k$ and $c$ represent the number of training epochs, the length of recommendation list and the number of queries, respectively.}
\label{tab:pre}
\end{table*}

Our idea is that limiting the access to the target RS data does not mean that the attacker cannot leverage a large volume of other public RS data to train the attack model.
We propose a new concept of \emph{Cross-system Attack}:
Thanks to the prosperous development of RS research, many real RS datasets are available and can be used for extracting knowledge and training the attacker to launch shilling attacks.
Along this direction, we design a \underline{P}ractical \underline{C}ross-system Shilling \underline{Attack} (\ours) framework that requires little information on the victim RS model and the target RS data. 
The contributions of this work are summarized as follows:
\begin{enumerate}
	\item We analyze the inadequacy of existing shilling attack methods and propose the concept of cross-system attack for designing a practical shilling attack model.

	\item We design \ours for shilling attacks. \ours is trained to capture graph topology knowledge from public RS data in a self-supervised manner. Then, it is fine-tuned on a small portion of target data that is readily available to construct fake profiles. \ours has all the three nice properties discussed above.

	\item We conduct extensive experiments to demonstrate that \ours exceeds state-of-the-art methods w.r.t. attack power and attack invisibility. Even in an unfair comparison where other attack methods can access the complete target data, \ours with limited access to the target data still exhibits superior performance.  
\end{enumerate}

\section{Background}
\label{sec:pre}

Shilling attacks can achieve both push attacks (promote the target item) and nuke attacks (demote the target item).
Since attackers can easily reverse the goal setting to conduct each attack~\citep{abs-2206-11433}, we consider push attacks in the sequel for simplicity.
In this paper, the \emph{source data} and the \emph{target data} refer to the RS data used for training the attack model and the data in the victim RS, respectively.

\subsection{Related Work}
\label{sec:related}

Early works of shilling attacks rely on heuristics~\citep{GunesKBP14}. 
Recent works~\cite{DeldjooNM21} mostly adopt the idea of adversarial attack~\citep{YuanHZL19}, and they can be categorized into four groups.

\textbf{Optimization methods} study how to model shilling attacks as an optimization task and then use optimization strategies to solve it.
\citet{LiWSV16} assumes the victim RS adopts matrix factorization (MF) and they propose methods PGA and SGLD that directly add the attack goal into the objective of MF.
RevAdv proposed by~\citet{TangWW20} and RAPU proposed by~\citet{ZhangTLSYZG21} model shilling attacks as a bi-level optimization problem.

\textbf{GAN-based methods} adopt Generative Adversarial Network (GAN)~\citep{GoodfellowPMXWOCB14} to construct fake user profiles. The generator models the data distribution of real users and generates real-like data, while the discriminator is responsible for identifying the generated fake users. 
Along this direction, a large number of methods have sprung up: TrialAttack~\citep{WuLGZC21}, Leg-UP~\citep{abs-2206-11433}, DCGAN~\citep{Christakopoulou19}, AUSH~\citep{LinC0XLY20}, RecUP~\citep{ZhangCZWL21}, to name a few. 

\textbf{RL-based methods} query the RS to get feedback on the attack. Then, Reinforcement Learning (RL)~\citep{KaelblingLM96} is used to adjust the attack. Representative works include PoisonRec~\citep{SongLHWLLG20}, LOKI~\citep{ZhangLD020} and CopyAttack~\citep{FanDZ0LWT021}. 

\textbf{KD-based methods} leverage Knowledge Distillation (KD)~\citep{GouYMT21} to narrow the gap between the surrogate RS and the victim RS. The surrogate RS is used to mimic the victim RS when the prior knowledge is not available. 
Model Extraction Attack proposed by \citet{YueHZM21} falls in this category.

\subsection{Analysis of Existing Works}
\label{sec:analysis}

We review existing shilling attack approaches and summarize their \emph{characteristics} in Tab.~\ref{tab:pre}:
\begin{itemize}
	\item \textbf{Data Knowledge}: Some methods assume the complete/partial target data is exposed to attackers. A practical attack method should use as less target data as possible.
	\item \textbf{RS Parameter Knowledge}: Some methods require the knowledge of the learned parameters of the victim RS. Such information is typically not available. 
	\item \textbf{RS Architecture Knowledge}: Some methods require the knowledge of the architecture of the victim RS. Such information is typically not available. 
	\item \textbf{Train with a surrogate RS}: Use a surrogate RS to train the attacker to avoid the prior knowledge of the victim RS.
	\item \textbf{Require multiple queries}: Query the victim RS multiple times and adjust fake  profiles according to the feedback.
	\item \textbf{Cross-domain Attack}: Use the information in one RS domain to attack another RS domain, e.g., train on the book data in Amazon RS and then attack video items in Amazon RS. Source and target domains share users and/or items.
	\item \textbf{Cross-system Attack}: Use the information in one RS to attack another RS, e.g., train on the Yelp RS and then attack Amazon RS. Source RS and target RS may not share users and/or items.
\end{itemize}

\begin{figure*}[t]
\centering
\includegraphics[width=1\linewidth]{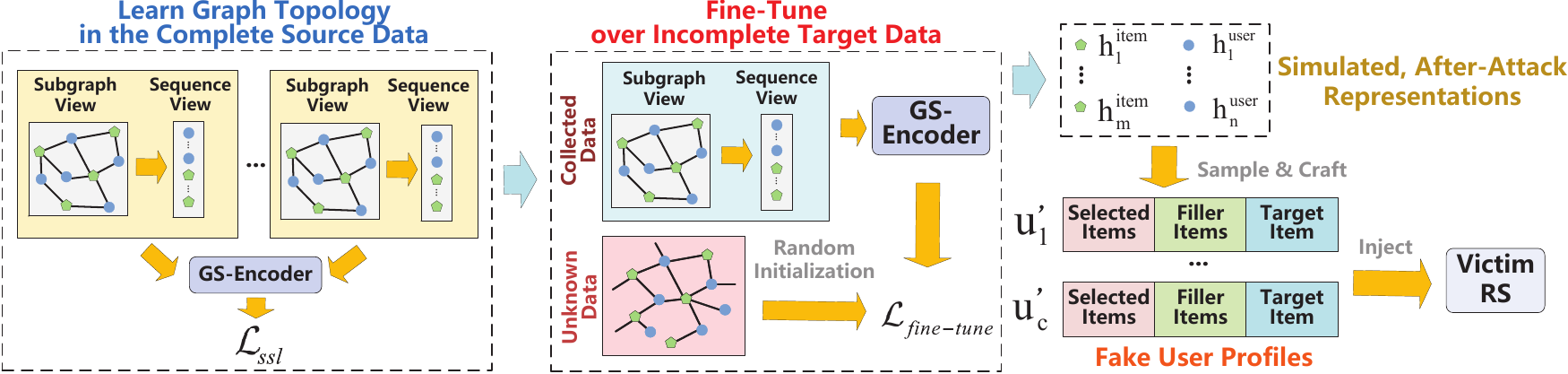}
\caption{Overview of \ours.}
\label{fig:framework}
\end{figure*}

From Tab.~\ref{tab:pre}, we can see that 
none of the existing methods have all the three properties illustrated in Sec.~\ref{sec:intro}.
In other words, \emph{there is still no real practical shilling attack method}.
Particularly, we do not find any method that has Property 2 and can achieve cross-system attack. 
Property 2 partially manifests in CopyAttack.
But CopyAttack assumes the source data and the target data share items and it is only able to achieve cross-domain attack.
Model Extraction Attack considers a data-free setting and uses limited queries ($c$ queries) to close the gap between the surrogate RS and the victim RS.
But its idea only works on sequential RS and the number of required queries is hard to pre-defined.

In summary, based on Tab.~\ref{tab:pre}, we can conclude that our method \ours (illustrated in the next section) has all of the three properties that a practical shilling attack method should have. 
And it is able to achieve cross-system attack, a difficult but practical setting of real shilling attacks.

\section{Our Framework \ours}
\label{sec:method}
\subsection{Motivation and Overview of \ours}

Existing works on cross-domain and cross-system recommendations~\citep{ZhaoPXZLY13,ZhuWCLOW18} have verified that the knowledge learned from a source domain/system can help improve recommendation results in a target domain/system, i.e., the RS knowledge is \emph{transferable}.
This has inspired us in designing \ours.
We believe it is possible to have an attack model that captures RS knowledge from the source data and can be transferred to attack the victim RS.
Fig.~\ref{fig:framework} provides an overview of \ours:

\begin{enumerate}
\item Firstly, we construct a bipartite user-item graph on the source data where each user-item edge indicates the existence of the corresponding user-item interaction.
\item After that, \ours trains a graph structure encoder (GS-Encoder) to capture the structural patterns of the source data in a self-supervised manner.
\item Then, \ours feeds a small portion of the public target data (e.g., a popular item and some people who bought it) into GS-Encoder and fine-tunes to get simulated representations after a successful attack.
\item Finally, based on the simulated representations, \ours searches for possible co-visit items of the target item that affect the possibility of recommending the target item and fills them in fake user profiles. Fake user profiles are injected into the victim RS to start the attack.
\end{enumerate}

\ours does not assume that the entity-correspondences across different domains/systems exist (i.e., a user or an item exists in both source and target data) even if they indeed exist.
This way, \ours does not require additional prior knowledge.
Therefore, in step 2, \ours is designed to only capture the structural patterns of the source data without knowing real identities of each node.
To endow \ours with the attack power, when constructing fake user profiles in step 4, we adopt the idea that items that can affect whether the target item is recommended are likely to have been interacted by some users together with the target item. 
This idea is called as \emph{co-visitation attack} and has been verified in existing shilling attack methods~\citep{YangGC17}.
Compared to existing methods, \ours requires \emph{much less} information.
It only needs a subgraph of a popular item, numbers of users/items in the target data and users who  interacted with the target item before.

\subsection{Learn from Graph Topology in the Source}

Without the knowledge of explicit entity-correspondences between the source data and the target data, we can not directly leverage historical records in the source data to encode users and items into representations that can be later used in attacking the target data.
Nevertheless, previous studies have shown that topologies of user-item bipartite graphs from different RS data share some common properties~\citep{HuangZ11}.
We can construct a user-item bipartite graph from the source data and train a graph structure encoder (GS-Encoder) to capture graph topological properties that are shared among different RS domains/systems.

Graph Neural Network (GNN) is the prevalent neural network used for modeling graph data. 
We use GNN as the backbone of GS-Encoder to capture the intrinsic and transferable properties from interaction data in the bipartite graph.
However, most feedback provided by users is implicit (e.g., clicks and views) rather than explicit (e.g., ratings).
Hence, the observed interactions often contain noise that may not indicate real user preferences. 
Neighborhood aggregation schemes in GNN may amplify the influence of interactions on representation learning, making learning more susceptible to interaction noise.

To alleviate the negative effect of noise, we introduce \emph{contrastive learning}, a type of self-supervised learning (SSL)~\citep{LiuZGNWZT21}, into GS-Encoder.
SSL constructs supervision signals from the correlation within the input data, avoiding requiring implicit labels.
Through contrasting samples, GS-Encoder learns to move similar sample pairs to stay close to each other while dissimilar ones are far apart.
We adopt the idea of \emph{multi-view learning}~\citep{abs-2103-00111} when designing the self-supervised contrastive learning task.
We model the node neighborhood as both a subgraph and a sequence, which helps GS-Encoder better capture the topological properties.

\subsubsection{Multi-view Data Augmentation}
To model node neighborhoods, we first sample paths by random walks and expand a single node $j$ in the user-item bipartite graph into its local structure as the \emph{subgraph view} $g_j$.
We use the random walk with a restart process~\citep{TongFP06}: (1) a random walk starts from a node $j$ in the bipartite graph, and (2) it randomly transmits to a neighbor with a probability $\alpha$ or returns to $j$ with a probability $1- \alpha$ in each step.
Note that we re-construct subgraph views in each training epoch.

For each node, its 1-hop neighbors describe \emph{user-item interaction patterns}. 
The 2-hop neighbors exhibit \emph{co-visitation patterns} (i.e., users who have interacted with the same item or items which have been interacted by the same user), which are important in shilling attacks~\citep{YangGC17}.
We propose to construct the \emph{sequence view} to capture the above two types of patterns better.
With the node $j$ as the center, we sort by the ID of its 1-hop nodes and its 2-hop nodes in turn to construct the sequence view $s_j$ of $j$.
The difference between the sequence view and the subgraph view is that the sequence view directly separates two data patterns while the subgraph view mixes two patterns up. Using the sequence view emphasizes learning two patterns individually while using the subgraph view learns them as a whole.

\subsubsection{Multi-View Contrastive Learning}  
Contrastive learning aims to maximize the similarity between positive samples while minimizing the similarity between negative samples.  
A suitable contrast task will facilitate capturing topological properties from the source data. 
Unlike most contrastive learning methods that only focus on contrasting positive and negative samples in one view, we deploy a multi-view contrast mechanism when designing GS-Encoder so that GS-Encoder can benefit from more supervision signals.

The subgraph view of each node is passed to a GNN encoder in GS-Encoder. 
We adopt GIN~\citep{XuHLJ19} as the GNN encoder, though other GNNs can be adopted.
The GNN encoder updates node representations as follows:
\begin{equation}
	\begin{aligned}
		\mathbf{h}_v^{(b)} = \text{MLP}^{(b)}\big((1 + \epsilon^{(b)}) \cdot \mathbf{h}_v^{(b-1)} 
			+ \Sigma_{u \in \mathcal{N}(v)}\mathbf{h}_u^{(b-1)}\big),
	\end{aligned}
\end{equation}
where $\mathbf{h}_v^{(b)}$ is the representation of node $v$ at the $b$-th GNN layer,  $\mathcal{N}(v)$ is the set of 1-hop nodes to $v$ and $\text{MLP}(\cdot)$ indicates multi-layer perceptron. 
We use eigenvectors of the normalized graph Laplacian of the subgraph to initialize $\mathbf{h}^{(0)}$ of each node in the subgraph~\citep{QiuCDZYDWT20}.
For a node $j$, its representation $\mathbf{h}_j^g$, from the subgraph view, is the concatenation of the aggregation of its neighborhood's representations generated in all GNN layers: 
\begin{equation}
	\begin{aligned}
		\mathbf{h}_j^g = \text{Concat}\big(\text{Readout}(\{\mathbf{h}_v^{(b)}|v \in \mathcal{V}_j \})\,|\,b=0,1,...,\hat{b}\big),
	\end{aligned}
\end{equation}
where $\text{Concat}(\cdot)$ denotes the concatenate operation, the $\text{Readout}(\cdot)$ function aggregates representations of nodes in the subgraph of $j$ from each iteration, and $\hat{b}$ is the number of GIN layers.

The sequence view of each node $j$ is passed to a LSTM and we use the last hidden state $\mathbf{h}^{s}_j$ as the representation from the sequence view.

$\mathbf{h}^g$ and $\mathbf{h}^s$ are further fed to a fully connected feedforward neural network to map them to the same latent space:
\begin{equation}
	\label{eq:mapping}
	\begin{aligned}
		\hat{\mathbf{h}}_j^g = \textbf{W}_2 \cdot \sigma(\textbf{W}_1 \mathbf{h}_j^g + \textbf{b}_1) + \textbf{b}_2, \\
		\hat{\mathbf{h}}_j^{s} = \textbf{W}_2 \cdot \sigma(\textbf{W}_1 \mathbf{h}_j^{s} + \textbf{b}_1) + \textbf{b}_2,
	\end{aligned}
\end{equation}
where {$\textbf{W}_1, \textbf{W}_2, \textbf{b}_1, \textbf{b}_2$} are learnable weights, and $\sigma(\cdot)$ indicates the sigmoid function.

In contrastive learning, we need to define positive and negative samples.  
For each node $j$, we define its positive sample $\mathit{pos}_j$ and negative samples $\mathit{neg}_j$ as the subgraph obtained by random walks starting from $j$ and its corresponding sequence view, and the subgraphs obtained by random walks starting from other nodes and their corresponding sequence views, respectively.
Note that, to improve efficiency and avoid processing too many negative subgraphs/sequences, we use subgraph/sequence views of other nodes in the same batch as negative samples.

The contrastive loss under the subgraph view is:
\begin{equation}
	\begin{aligned}
		\mathcal{L}_j^{g} = -log 
		\frac{exp\big(sim(\hat{\mathbf{h}}_j^g, \hat{\mathbf{h}}_{\mathit{pos}_j}^{s})/\tau\big)}
		{\Sigma_{l \in \mathit{neg}_j}exp\big(sim(\hat{\mathbf{h}}_j^g, \hat{\mathbf{h}}_l^{s})/\tau\big)},
	\end{aligned}
\end{equation}
where $\text{sim}(\cdot)$ denotes the cosine similarity and $\tau$ denotes a temperature parameter. 
The contrastive loss under the sequence view is defined similarly:
 \begin{equation}
	\begin{aligned}
		\mathcal{L}_j^{s} = -log 
		\frac{exp\big(sim(\hat{\mathbf{h}}_j^{s}, \hat{\mathbf{h}}_{\mathit{pos}_j}^{g})/\tau\big)}
		{\Sigma_{l \in \mathit{neg}_j}exp\big(sim(\hat{\mathbf{h}}_j^{s}, \hat{\mathbf{h}}_l^{g})/\tau\big)}.
	\end{aligned}
\end{equation}

The overall multi-view contrastive objective of GS-Encoder is as follows:
\begin{equation}
	\begin{aligned}
		\mathcal{L}_{ssl} = \frac{1}{n} \Sigma_{j \in \mathcal{I}}
		(\lambda_g \cdot \mathcal{L}_j^{g} + \lambda_{s} \cdot \mathcal{L}_j^{s}),
	\end{aligned}
\end{equation}
where $\lambda_g$ and $\lambda_{s}$ are hyper-parameters to balance two views, $\mathcal{I}$ is the item set, and $n$ is the number of items.

\subsection{Craft Fake User Profiles in the Victim RS}

After pre-training the GS-Encoder, the next step is to construct a few fake user profiles and inject them into the victim RS to pollute the target data.
Our construction method of fake users is based on three design principles from the literature:
\begin{itemize}
\item \textbf{Principle 1:} Item-based RS is designed to recommend items similar to past items in the target user's profile~\citep{SarwarKKR01}.  

\item \textbf{Principle 2:} User-based RS is designed to recommend items interacted by similar users of the target user~\citep{HerlockerKR00}.  

\item \textbf{Principle 3:} According to the idea of co-visitation attack~\citep{YangGC17}, the co-visit items of the target item (i.e., the 2-hop neighbors of the target item in the bipartite graph) can affect whether the target item can be recommended.  
\end{itemize}

Based on the above principles, the goals of our construction method are:
\begin{itemize}
\item \textbf{Goal 1:} Based on Principle 1, our goal is to affect the victim RS so that the representation of target item is as similar as possible to representations of the rest of the items. This way, the possibility of recommending the target item can increase. We hope that our attack can achieve the following objective: for any item $i$, $sim(\mathbf{h}^{item}_i, \mathbf{h}^{item}_t)>sim(\mathbf{h}^{item}_i, \mathbf{h}^{item}_j)$, where $t$ is the target item, $j$ denotes any other item, and $sim(\cdot)$ denotes a measure of similarity between items (e.g., cosine similarity), and $\mathbf{h}^{item}_i$ is the representation of item $i$ in the victim RS.

\item \textbf{Goal 2:} Based on Principle 2, our goal is to affect the victim RS so that representations of users who have interacted with the target item are as similar as possible to representations of other users. This way, the possibility of recommending the target item can increase. We hope that, for any user $u$, $sim(\mathbf{h}^{user}_u, \mathbf{h}^{user}_r)>sim(\mathbf{h}^{user}_u, \mathbf{h}^{user}_e)$, where $r \in \mathcal{N}(t)$ denotes the user who has interacted with the target item $t$, $e \notin \mathcal{N}(t)$ is any other user, and $\mathbf{h}^{user}_u$ is the representation of user $u$ in the victim RS.

\item \textbf{Goal 3:} Based on Principle 3, our goal is to find possible co-visit items of the target item after a successful attack and fill them in the fake user profiles.
\end{itemize}

However, the above goals are challenging without knowing the details of the victim RS and the target data, i.e., we do not know $\mathbf{h}^{item}$ and $\mathbf{h}^{user}$ in the victim RS. 
GS-Encoder, which captures the transferable knowledge from the more informative source data, can help us accomplish this task:
\begin{itemize}
\item \textbf{Step 1:} Use the pre-trained GS-Encoder to generate node representations based on topological information of the incomplete target data. 

\item \textbf{Step 2:} Fine-tune and get \emph{simulated} representations \emph{after the successful attack} (Goal 1 and Goal 2). 

\item \textbf{Step 3:}  Based on the simulated, after-attack representations, we search for possible co-visit items of the target items and craft the fake user profiles (Goal 3).
\end{itemize}

\begin{figure}[t]
\centering
\includegraphics[width=1\columnwidth]{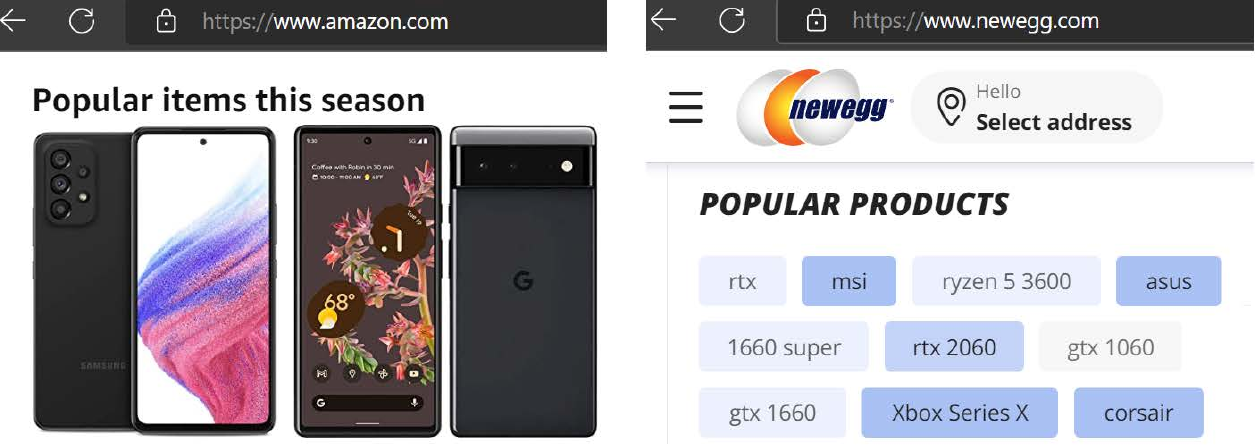}
\caption{Popular items of Amazon and Newegg.}
\label{fig:examples}
\end{figure}

Considering that we cannot access the complete target data, we collect a very small portion of the target data that can be publicly accessed.
One example is the popular item in the victim RS, some normal users who have bought them and the 2-hop items to the popular item. 
Such information is typically available.
For instance, Amazon provides ``Popular Items This Season'' and Newsegg provides ``Popular Products'' on their homepages, as shown in Fig.~\ref{fig:examples}, 
and information of their buyers can be found by clicking the popular item.
The buyer's homepage may provide information of some items that he/she bought before.
Therefore, starting from one popular item, we collect users/items in its 2-hop subgraph via random walks without restart. 
But we limit the total number of nodes to be lower than $p$ percentage of the target data to keep a low level of knowledge.
In addition to the collected subgraph of a popular item, we collect a user set $\mathcal{M}(t)$ containing users who have interacted with the target item $t$.

Based on the collected data from the target data, we construct a small subgraph centering on the popular item, and feed it into the pre-trained GS-Encoder to generate initial representations of users/items in the subgraph:
\begin{equation}
	\begin{aligned}
		\mathbf{h}_j = \eta_g \cdot \hat{\mathbf{h}}_j^g + \eta_s \cdot \hat{\mathbf{h}}_j^{s},
	\end{aligned}
\end{equation}
where $\eta_g$ and $\eta_s$ are hyper-parameters that balance the effects of the two views, $\mathbf{h}_j$ is the fused representation of node $j$, and $\hat{\mathbf{h}}_j^g$ and $\hat{\mathbf{h}}_j^s$ are subgraph-view representation and sequence-view representation of node $j$ generated by the pre-trained GS-Encoder as shown in Eq.~\ref{eq:mapping}, respectively.

For most users and items in the target data that are not collected, we assume that we know the numbers of users ($m$) and items ($n$) in the victim RS, and initialize their representations from a normal distribution $\mathcal{N}(0,0.1)$.
This is a reasonable assumption as many RS websites reveal exact numbers or the order of magnitude of users/items.
Including users and items that are not in the collected small subgraph makes it possible to generate fake user profiles with items not in the collected subgraph.

We continue to fine-tune over the above collected data with the following objective for Goal 1 to simulate representations after a successful attack:
\begin{equation}
	\begin{aligned}
		\mathcal{L}^{item} = -log \Sigma_{i=1}^{n} 
		\frac{sim(\mathbf{h}_i^{item}, \mathbf{h}_t^{item})}{\Sigma_{j\neq t} sim(\mathbf{h}_i^{item}, \mathbf{h}_j^{item})}. 
	\end{aligned}
\end{equation}

Similarly, for Goal 2, we fine-tune with the following objective to simulate representations after a successful attack:
\begin{equation}
	\begin{aligned}
		\mathcal{L}^{user} = -log \Sigma_{u=1}^{m} 
		\frac{\Sigma_{g\in \mathcal{M}(t)} sim(\mathbf{h}^{user}_u, \mathbf{h}^{user}_g)}
		{\Sigma_{j \notin \mathcal{M}(t)} sim(\mathbf{h}^{user}_u, \mathbf{h}^{user}_j)}. 
	\end{aligned}
\end{equation}

The overall objective is as follows:
\begin{equation}
	\begin{aligned}
		\mathcal{L}_{fine-tune}  =\mu_{item}\cdot \mathcal{L}^{item} + \mu_{user}\cdot \mathcal{L}^{user},
	\end{aligned}
\end{equation}
where $\mu_{item}$ and $\mu_{user}$ are hyper-parameters that balance the effects of the two loss functions.

After fine-tuning, we have new representations of users and items that simulate the representations in the victim RS after a successful attack. 
Then, we search for possible co-visit items of the target item based on the simulated representations.
Similar to other shilling attack methods~\citep{LinC0XLY20,abs-2206-11433}, the fake user profile in \ours contains three parts: selected items, filler items and the target item.
We estimate the potential interest of all users in the target item $t$ after the attack by the inner product of representations, and sample $z$ users according to the probability:
\begin{equation}
Pro(u|t) = \frac{\mathbf{h}^{item}_t \cdot \mathbf{h}^{user}_u}{\Sigma_{j=1}^{m} \mathbf{h}^{item}_t \cdot \mathbf{h}^{user}_j}.
\end{equation}
Common items existing in these $z$ profiles are chosen as the selected items.
Because popular items are always more accessible than others and appear in many normal users' profiles, we randomly sample $y$ popular items from the collected subgraph according to their degrees as filler items to enhance the invisibility of \ours.
For each fake profile, the above crafting process is conducted independently.

\section{Experiments}
\label{sec:exp}

\begin{table}[t]
	\centering
	\scalebox{0.9}{
		\begin{tabular}{ccccc}
		\hline
		Dataset    & \#Users & \#Items & \#Interactions & Sparsity \\ \hline
		FilmTrust  & 780     & 721     & 28,799         & 94.88\%  \\
		Automotive & 2,928   & 1,835   & 20,473         & 99.62\%  \\
		T \& HI    & 1,208   & 8,491   & 28,396         & 99.72\%  \\
		Yelp       & 2,762   & 10,477  & 119,237        & 99.59\%  \\ \hline
		\end{tabular}
	}
	\caption{Statistics of datasets}
	\label{tab:statistics}
\end{table}

\begin{table*}[t]
	\centering
	\resizebox{0.95\linewidth}{!}{
		\begin{tabular}{c|c|ccccccccc}
		\hline
		\multirow{2}{*}{Dataset}    & \multirow{2}{*}{Victim RS} & \multicolumn{9}{c}{Attack Method (HR@50)}                                                                                                     \\ \cline{3-11} 
		                            &                            & RevAdv & TrialAttack     & Leg-UP          & AUSH   & Bandwagon       & Random          & Segment         & PC-Attack$^*$   & PC-Attack       \\ \hline
		\multirow{7}{*}{Automotive} & CDAE                       & 0.1643 & \textbf{0.2504} & 0.2237          & 0.1949 & 0.2114          & 0.2247          & 0.1949          & 0.1348          & 0.1693          \\
		                            & ItemAE                     & 0.2916 & 0.3128          & 0.3105          & 0.2926 & 0.3232          & 0.3183          & 0.2926          & \textbf{0.4908} & 0.3787          \\
		                            & LightGCN                   & 0.1002 & 0.1228          & 0.1149          & 0.1361 & 0.1462          & 0.1222          & 0.1361          & 0.1406          & \textbf{0.1716} \\
		                            & NCF                        & 0.7012 & 0.7744          & 0.7712          & 0.7359 & 0.7689          & 0.7462          & 0.7359          & \textbf{0.8574} & 0.6484          \\
		                            & NGCF                       & 0.0866 & 0.0960          & 0.1398          & 0.1416 & 0.1362          & 0.1387          & \textbf{0.1416} & 0.0864          & 0.1056          \\
		                            & VAE                        & 0.0842 & 0.0841          & 0.1168          & 0.1196 & 0.0960          & 0.0965          & 0.1196          & 0.1238          & \textbf{0.1559} \\
		                            & WRMF                       & 0.9243 & 0.2941          & 0.3561          & 0.4386 & 0.3769          & 0.3069          & 0.4386          & \textbf{0.9268} & 0.9213          \\ \hline
		\multirow{7}{*}{FilmTrust}  & CDAE                       & 0.4587 & 0.6270          & 0.6190          & 0.5342 & 0.4837          & 0.5954          & 0.5342          & 0.7505          & \textbf{0.7810} \\
		                            & ItemAE                     & 0.6429 & 0.4721          & 0.5965          & 0.5501 & 0.5253          & 0.4807          & 0.5501          & 0.9534          & \textbf{0.9544} \\
		                            & LightGCN                   & 0.8522 & 0.8362          & 0.8820          & 0.8594 & 0.8271          & 0.8169          & 0.8594          & \textbf{0.8949} & 0.8517          \\
		                            & NCF                        & 0.9319 & 0.9108          & 0.9521          & 0.8846 & 0.8855          & 0.8929          & 0.8846          & 0.9266          & \textbf{0.9543} \\
		                            & NGCF                       & 0.9015 & 0.9123          & \textbf{0.9207} & 0.9072 & 0.9079          & 0.9091          & 0.9072          & 0.9037          & 0.9101          \\
		                            & VAE                        & 0.9713 & 0.9742          & \textbf{0.9749} & 0.9724 & 0.9721          & 0.9726          & 0.9724          & 0.9689          & 0.9730          \\
		                            & WRMF                       & 0.5143 & 0.4171          & 0.4732          & 0.4976 & 0.4873          & 0.4667          & 0.4976          & \textbf{0.5706} & 0.4935          \\ \hline
		\multirow{7}{*}{T \& HI}    & CDAE                       & 0.1126 & 0.3186          & 0.3929          & 0.3409 & 0.3173          & \textbf{0.4156} & 0.3409          & 0.3449          & 0.2279          \\
		                            & ItemAE                     & 0.1074 & 0.2755          & 0.2677          & 0.1433 & 0.1857          & 0.2335          & 0.1433          & \textbf{0.3324} & 0.1487          \\
		                            & LightGCN                   & 0.0028 & 0.0376          & 0.0383          & 0.0531 & 0.1603          & 0.0273          & 0.0531          & 0.0456          & \textbf{0.2064} \\
		                            & NCF                        & 0.5421 & 0.6895          & 0.8508          & 0.8038 & \textbf{0.9017} & 0.8827          & 0.8038          & 0.7749          & 0.1988          \\
		                            & NGCF                       & 0.0177 & 0.0739          & 0.1018          & 0.0903 & 0.1016          & 0.0927          & 0.0903          & \textbf{0.1064} & 0.0705          \\
		                            & VAE                        & 0.3530 & 0.9975          & 0.9993          & 0.9995 & 0.9991          & \textbf{0.9996} & 0.9995          & 0.9916          & 0.9669          \\
		                            & WRMF                       & 0.0406 & 0.0697          & 0.0495          & 0.0448 & 0.0530          & 0.0460          & 0.0448          & \textbf{0.0868} & 0.0743          \\ \hline
		\end{tabular}
	}
	\caption{Attack performance (HR@50) of different attack methods against different victim RS models. \oursstar indicates that the complete target data is used. Best results are shown in bold.}
	\label{tab:main_exp}
\end{table*}

\begin{table*}[t]
  \begin{minipage}[b]{0.42\linewidth}
	\centering
	\includegraphics[width=0.88\linewidth]{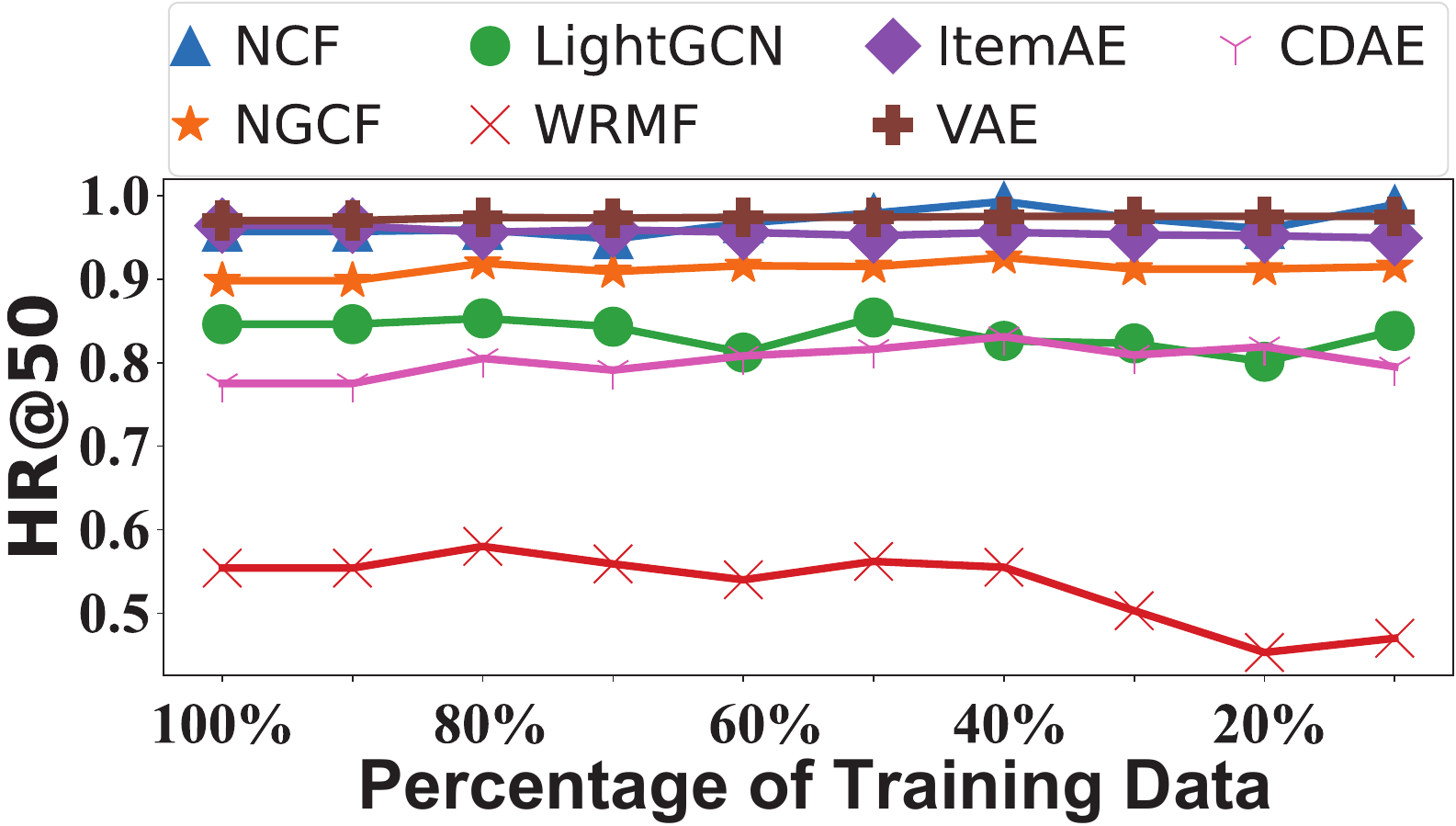}
	\captionof{figure}{Impacts of Accessible Target Data ($p$).}
	\label{fig:exp-data}
  \end{minipage}
  \hfill
  \begin{varwidth}[b]{0.56\linewidth}
    \centering
    \resizebox{1\linewidth}{!}{
		\begin{tabular}{ccccccc}
			\hline
			\multirow{3}{*}{\begin{tabular}[c]{@{}c@{}}Victim\\ RS\end{tabular}} & \multicolumn{6}{c}{Source Data}                                                                     \\ \cline{2-7} 
			                                                                     & \multicolumn{2}{c}{Automotive}  & \multicolumn{2}{c}{T \& HI}     & \multicolumn{2}{c}{Yelp}        \\
			                                                                     & HR@50          & NDCG@50        & HR@50          & NDCG@50        & HR@50          & NDCG@50        \\ \hline
			CDAE                                                                 & 0.773          & 0.198          & \textbf{0.793} & \textbf{0.199} & 0.781          & 0.200          \\
			ItemAE                                                               & 0.953          & 0.257          & 0.949          & 0.255          & \textbf{0.954} & \textbf{0.257} \\
			LightGCN                                                             & 0.817          & 0.219          & 0.823          & 0.221          & \textbf{0.852} & \textbf{0.236} \\
			NCF                                                                  & 0.907          & 0.247          & 0.927          & 0.254          & \textbf{0.954} & \textbf{0.256} \\
			NGCF                                                                 & 0.895          & 0.243          & 0.909          & 0.242          & \textbf{0.910} & \textbf{0.247} \\
			VAE                                                                  & 0.971          & 0.259          & 0.970          & 0.258          & \textbf{0.973} & \textbf{0.259} \\
			WRMF                                                                 & \textbf{0.508} & \textbf{0.139} & 0.475          & 0.131          & 0.494          & 0.136          \\ \hline
		\end{tabular}
	}
	\caption{Results of using different source datasets (the target dataset is FilmTrust). Best results are shown in bold.}
	\label{tab:src-to-filmtrust}
  \end{varwidth}
\end{table*}

\begin{table*}[t]
  \begin{minipage}[b]{0.42\linewidth}
	\centering
    \includegraphics[width=0.8\columnwidth]{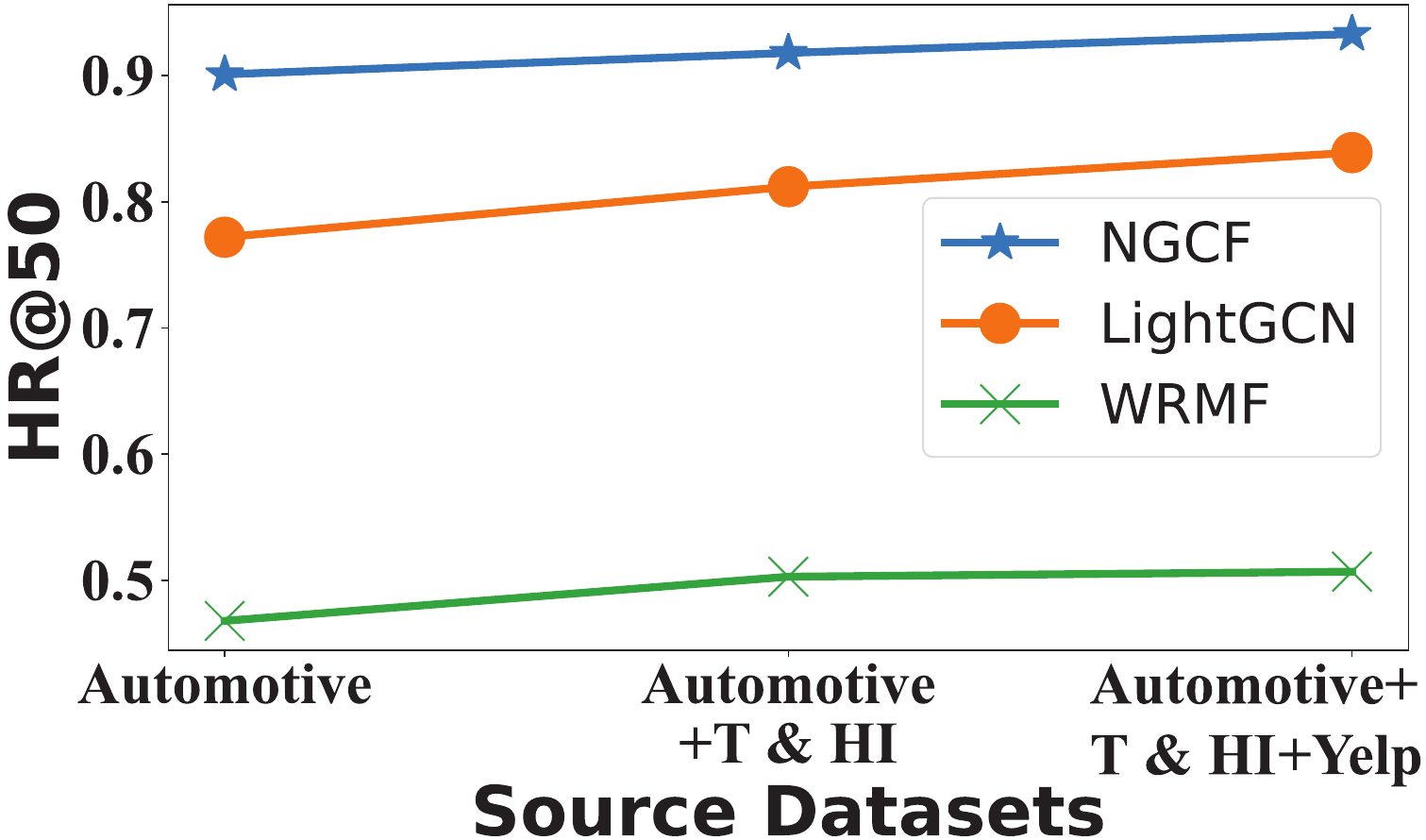}
	\captionof{figure}{Learn graph topology from multiple source datasets and then attack victim RS on FilmTrust.}
	\label{fig:cross-doamin}
  \end{minipage}
  \hfill
  \begin{varwidth}[b]{0.56\linewidth}
    \centering
	\resizebox{0.98\linewidth}{!}{
		\begin{tabular}{ccccccc}
		\hline
		\multirow{3}{*}{\begin{tabular}[c]{@{}c@{}}Victim\\ RS\end{tabular}} & \multicolumn{6}{c}{Target Data}                                                                           \\ \cline{2-7} 
		                                                                     & \multicolumn{2}{c}{FilmTrust}     & \multicolumn{2}{c}{Automotive}    & \multicolumn{2}{c}{T \& HI}       \\
		                                                                     & Precison        & Recall          & Precison        & Recall          & Precison        & Recall          \\ \hline
		RevAdv                                                               & 0.0713          & 0.0778          & 0.0080          & 0.0089          & \textbf{0.0180} & \textbf{0.0200} \\
		TrialAttack                                                          & 0.1603          & 0.1927          & 0.1114          & 0.1038          & 0.0200          & 0.0256          \\
		Leg-UP                                                               & 0.2497          & 0.2711          & 0.0000          & 0.0000          & 0.0400          & 0.0444          \\
		AUSH                                                                 & 0.2429          & 0.2556          & 0.0380          & 0.0422          & 0.0742          & 0.0822          \\
		Bandwagon                                                            & 0.2371          & 0.2489          & 0.0220          & 0.0244          & 0.0762          & 0.0845          \\
		Random                                                               & 0.2340          & 0.2467          & 0.0220          & 0.0244          & 0.0441          & 0.0489          \\
		Segment                                                              & 0.2602          & 0.2733          & 0.0280          & 0.0311          & 0.0662          & 0.0733          \\
		PC-Attack                                                            & \textbf{0.0280} & \textbf{0.0311} & \textbf{0.0000} & \textbf{0.0000} & 0.0240          & 0.0266          \\ \hline
		\end{tabular}
    }
   \caption{Detection performance on different attack methods. Best results are shown in bold.}
   \label{tab:det}
  \end{varwidth}
\end{table*}

\subsection{Experimental Settings} 

\subsubsection{Datasets}
We use four public datasets\footnote{\url{https://github.com/XMUDM/ShillingAttack} \label{footnote:legup}} widely adopted in previous works on shilling attacks~\citep{LinC0XLY20,abs-2206-11433}, including FilmTrust, Yelp and two other Amazon datasets Automotive, and Tools \& Home Improvement (T \& HI).
Target items for testing attacks are included in the datasets.
Tab.~\ref{tab:statistics} illustrates the statistics of the data.
Default training/test split is used for training and tuning surrogate RS models (if baselines require a surrogate RS) and victim RS models.
By default, we train \ours to learn graph topology from the complete Yelp dataset since Yelp is the largest dataset. 
Then, we test it on attacking victim RS on the other three datasets.
Since some experiments require long-tail items, we define long-tail items as items with no more than three interactions.

\subsubsection{Shilling Attack Baselines}
We use three classic attack methods\textsuperscript{\ref{footnote:legup}} Random Attack, Bandwagon Attack and Segment Attack~\citep{abs-2206-11433}, and four state-of-the-art shilling attack methods RevAdv\footnote{\url{https://github.com/graytowne/revisit_adv_rec}}~\citep{TangWW20}, TrialAttack\footnote{\url{https://github.com/Daftstone/TrialAttack}}~\citep{WuLGZC21}, Leg-UP\textsuperscript{\ref{footnote:legup}}~\citep{abs-2206-11433} and AUSH\textsuperscript{\ref{footnote:legup}}~\citep{LinC0XLY20} as baselines.

\subsubsection{Victim RS}
We conduct shilling attacks against various prevalent RS models: NCF~\citep{HeLZNHC17}, WRMF~\citep{HuKV08}, LightGCN~\citep{0001DWLZ020}, NGCF~\citep{Wang0WFC19}, VAE~\citep{LiangKHJ18}, CDAE~\citep{WuDZE16} and ItemAE~\citep{SedhainMSX15}.

\subsubsection{Hyper-parameters}
The hyper-parameters of attack baselines and victim RS are set as the original papers suggest and tuned to show the best results.
We set the number of fake profiles to 50 for all methods. 
This is roughly the population that can manifest the differences among attack models~\citep{BurkeMBW05}.
For \ours, we set training epochs to 32, batch size to 32, embedding size to 64 and learning rate to 0.005.
$z$ and $y$ used in crafting profiles are set to 50 and 10, respectively. 
The length of random walk is set to 64 and the restart probability $1-\alpha$ is 0.8.
The number of GIN layers $\hat{b}$ is 5.
Other hyper-parameters of \ours are selected through grid search and the chosen hyper-parameters are:
$\tau=0.07$, $\lambda_g=0.5$, $\lambda_s=0.5$, $\eta_g=0.5$, $\eta_s=0.5$, $\mu_{user}=0.5$, and $\mu_{item}=0.5$.
By default, we set $p=10\%$ when collecting target data.
Adam optimizer is adopted for optimization.

\subsubsection{Evaluation Metrics}
Hit Ratio (HR@k) and Normalized Discounted Cumulative Gain (NDCG@k) are used for evaluation.  
HR@k measures the average proportion of normal users whose top-k recommendation lists contain the target item after the attack.
NDCG@k measures the ranking of the target item after the attack.
For both metrics, we set k to 50.

\subsection{Overall Attack Performance}

Tab.~\ref{tab:main_exp} summarizes the overall attack performance of different attack methods.
We have the following observations:
\begin{enumerate}
	\item \oursstar and \ours together achieve the best results in most cases, showing the effectiveness of our designs. Some baselines may have better results in a few cases, but their attack performance is not robust.

	\item \ours achieves best results in more than 30\% cases. In other cases where \ours does not rank first, its performance is not far away from the best performance. Note that our goal (i.e., a practical attack) is to use as little information of the target data as possible. In the case of the unfair comparison, (i.e., baselines take the complete target data and \ours accesses at most 10\%), it is acceptable that \ours can have degraded performance in exchange for the feasibility of the attack. However, \ours shows promising results, showing the power of cross-system attack.
\end{enumerate}

\subsection{Impacts of Accessible Target Data ($p$)}

We evaluate the performance of \ours when $p$ changes (the default value is $p=10\%$ and \oursstar uses $p=100\%$). 
Fig.~\ref{fig:exp-data} illustrates the results on the FilmTrust dataset.
We can observe that, as $p$ decreases, the attack performance of \ours degrades gradually.
However, thanks to the knowledge learned from the source data, the decline of attack performance is not significant and \ours is robust when the percentage of accessible target data varies.

\subsection{Impacts of Source Data}

We conduct two experiments to check the impacts of source data on \ours:

\subsubsection{Impacts of Using Different Source Data} 
Tab.~\ref{tab:src-to-filmtrust} reports the results of \ours when using FilmTrust as the target data and the other three datasets are used as the source data. 
We can observe that using different source dataset does not affect the performance too much, which confirms that different RS data have common topological information of which the knowledge is transferable. 
However, the larger the source data is, the better \ours can capture the structural patterns. 
Hence, \ours shows best results when using Yelp (the largest dataset in our experiments) as the source data.

\subsubsection{Performance of Using Multiple Source Datasets} 
\ours can learn graph topology from multiple source datasets to benefit from the large volume of public RS data.
To illustrate the results of using multiple source dataset, we train \ours on different source datasets in the order of dataset size and then use it to attack WRMF, NGCF and LightGCN on the FilmTrust dataset.
As shown in Fig.~\ref{fig:cross-doamin}, as more source datasets are used, the performance of \ours gradually gets improved, showing that we can feed more public RS datasets to \ours and get even better attack performance.

\subsection{Attack Invisibility}

Next, we investigate the invisibility of \ours. 

\subsubsection{Attach Detection} 
We apply the state-of-the-art unsupervised attack detector~\citep{ZhangT0LCM15} on the fake profiles generated by different attack methods.
Tab.~\ref{tab:det} describes the accuracy and recall of the detector on different attack methods.
Lower precision and recall indicate that the attack method is less perceptible.
Based on the results, we find that the detection performance is highly data-dependent, and fake users are more easy to detect on denser datasets.  
For example, it is difficult for the detector to find fake users on Yelp. 
But it has relatively high precision and recall for detecting most attack methods (except \ours) on FilmTrust.
\ours generates almost undetectable fake users.  
In most cases, the detector performs the worst on \ours.  
On T \& HI, the detector does not has the lowest precision and recall for \ours, but the values are close to lowest ones.

\subsubsection{Fake User Distribution} 
\begin{figure}[!t]
	\centering
	\subfloat[Automotive]{
		\includegraphics[width=0.48\linewidth]{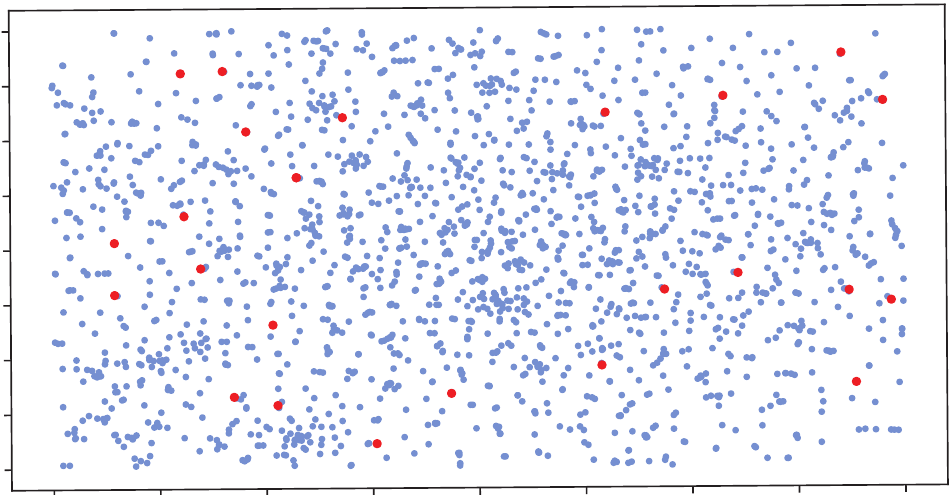}
	}
	\subfloat[FilmTrust]{
		\includegraphics[width=0.48\linewidth]{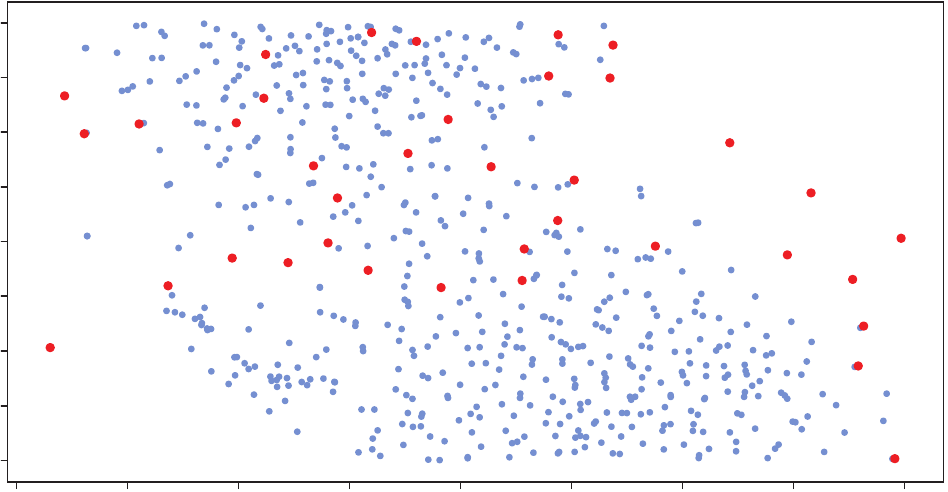}
	}
	\caption{Visualization of user profiles. Red nodes are fake profiles and blue nodes are real profiles.}
	\label{fig:vis}
\end{figure}

Using t-SNE~\citep{van2008visualizing}, Fig.~\ref{fig:vis} visualizes users' representations generated by WRMF after it is attacked by \ours on Automotive and FilmTrust.
We can observe that fake users profiles are scattered among real user profiles and it is hard for detectors to distinguish fake and real users, showing that \ours can launch virtually invisible attacks.

\subsection{Impacts of the Starting Node in Target Data} 

\begin{figure}[!t]
	\centering
	\includegraphics[width=0.8\linewidth]{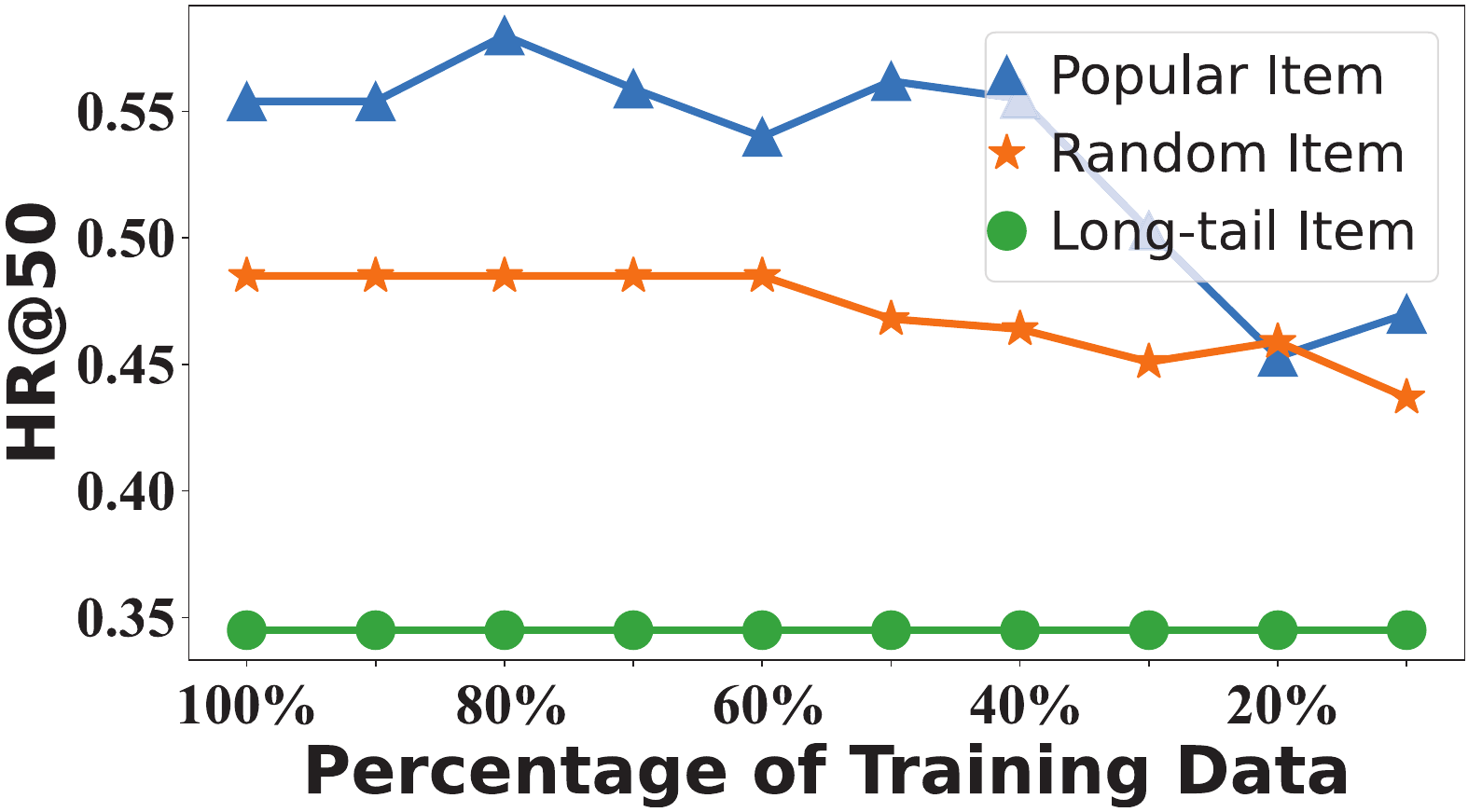}
	\caption{Impact of the starting item.}
	\label{fig:exp-data2}
\end{figure}

By default, we start with the most popular item to collect the target data and select no more than 10\% of user-item interaction records within the limit of only 2-hop neighbors.  
To evaluate the robustness of \ours, we compare the default setting with two other options: use an item sampled from long-tail items as the starting point and use a random sampled item as the starting point. 
Fig.~\ref{fig:exp-data2} compares the results of the default setting and two extra settings for attacking WRMF on FilmTrust.
We can see that the attack performance of starting with a randomly sampled item does not lag much behind starting with the most popular item, and is also robust w.r.t. different settings of total numbers of nodes in the collected target data.
Differently, starting from a sampled long-tail item results in a much worse performance. The reason is the long-tail item does not have many 2-hop neighbors and the collected data cannot help fine-tune GS-Encoder well on the target data.
The performance of starting from a long-tail item is very consistent when the limit of node numbers changes.
The reason is that long tail items have few neighbors and the number of the 2-hop neighbors of long tail items do not exceed 10\% of the total target data. Changing the limit does not actually change the number of the collected nodes.
In summary, starting with a popular item brings the best results. Considering that popular items are always more readily available, \ours uses the popular item as the default starting point to collect target data.

\subsection{Performance of Cross-domain Attack.}

\begin{table}[!t]
	\centering
	\scalebox{0.8}{
		\begin{tabular}{ccccc}
		\hline
		\multirow{2}{*}{\begin{tabular}[c]{@{}c@{}}Victim\\ RS\end{tabular}} & \multicolumn{2}{c}{T \& HI $\rightarrow$ Automotive} & \multicolumn{2}{c}{Automotive $\rightarrow$ T \& HI} \\ \cline{2-5} 
		                                                                     & HR@50                    & NDCG@50                   & HR@50                    & NDCG@50                   \\ \hline
		CDAE                                                                 & 0.035                    & 0.008                     & 0.222                    & 0.068                     \\
		ItemAE                                                               & 0.288                    & 0.096                     & 0.142                    & 0.080                     \\
		LightGCN                                                             & 0.160                    & 0.051                     & 0.238                    & 0.113                     \\
		NCF                                                                  & 0.366                    & 0.084                     & 0.079                    & 0.036                     \\
		NGCF                                                                 & 0.039                    & 0.009                     & 0.061                    & 0.018                     \\
		VAE                                                                  & 0.047                    & 0.011                     & 0.975                    & 0.565                     \\
		WRMF                                                                 & 0.902                    & 0.288                     & 0.073                    & 0.027                     \\ \hline
		\end{tabular}
	}
	\caption{Results of \ours for the cross-domain attack.}
	\label{tab:cross-domain}
\end{table}

\begin{table*}[!t]
	\centering
	\scalebox{0.8}{
		\begin{tabular}{ccc|ccc|ccc}
			\toprule
			\bm{$\eta_g:\eta_s$} & \textbf{HR@50} & \textbf{NDCG@50} & \bm{$\lambda_g:\lambda_s$} & \textbf{HR@50} & \textbf{NDCG@50} & \bm{$\mu_{item}:\mu_{user}$} & \textbf{HR@50} & \textbf{NDCG@50} \\  \cmidrule(l){1-9}  
			1:0 & 0.467 & 0.131 & 1:0 & 0.471 & 0.132 & 1:0 & 0.474 & 0.131 \\
			0:1 & 0.471 & 0.132 & 0:1 & 0.478 & 0.134 & 0:1 & 0.477 & 0.114 \\
			2:1 & 0.499 & 0.134 & 2:1 & 0.515 & 0.147 & 2:1 & 0.481 & 0.136 \\
			1:2 & 0.486 & 0.136 & 1:2 & 0.512 & 0.144 & 1:2 & 0.501 & 0.142 \\
			1:1 & 0.522 & 0.146 & 1:1 & 0.522 & 0.146 & 1:1 & 0.522 & 0.146 \\ \cmidrule(l){1-9}  
		\end{tabular}
	}
	\caption{Results using different hyper-parameters.}
	\label{tab:exp_para}
\end{table*}

\ours is designed for both cross-system attack and cross-domain attack. 
Experimental results in previous sections are for cross-system attack.
Next, we further report the performance of cross-domain attack using \ours.
Tab.~\ref{tab:cross-domain} provides the result of \ours for using T \& HI as the source and Automotive as the target, and using Automotive as the source and T \& HI as the target.
The two datasets contain data in different categories of Amazon.
From the result, we can observe that \ours achieves acceptable attack performance for cross-domain attack, but the performance is worse than cross-system attack reported in Tab.~3.
The reason is that the source dataset Yelp used in our default experiments for cross-system attack is much larger than T \& HI and Automotive used in the experiments for cross-domain attack.
\ours can better capture topological information from a larger source dataset.
Hence, it shows better results in cross-system attack than cross-domain attack.

\subsection{Impacts of Hyper-parameters and Ablation Study}

The three sets of balance hyper-parameters $\eta$, $\lambda$ and $\mu$ are used to balance the effects of representations from graph and sequence views, graph-view and sequence-view loss functions, and user and item loss functions, respectively.

Tab.~\ref{tab:exp_para} reports the performance of \ours when attacking WRMF using different balance hyper-parameters.
We can observe that the change of balance hyper-parameters affect the performance of \ours.
When setting equal values to all balance hyper-parameters, the attack performance is the best. 

Besides, the first two rows and the last row in Tab.~\ref{tab:exp_para} can be viewed as three ablation experiments: (1) only use graph-view representation, graph-view loss and item loss, (2) only use sequence-view representation, sequence-view loss and sequence loss, and (3) the default \ours that uses all parts.
From Tab.~\ref{tab:exp_para}, we can see that \ours performs best when all parts are present and removing any of them will degrade the attack performance.
Therefore, we can conclude that each component in \ours indeed contributes to its overall attack performance.

\section{Conclusion}
\label{sec:con}

In this paper, we study practical shilling attacks. 
We analyze the limitations of existing works and design a new framework \ours that transfers the knowledge to attack the victim RS on incomplete target dataset.
Experimental results demonstrate the superiority of \ours.
In the future, we plan to introduce more self-supervised learning tasks so that \ours can get more supervision signals and better capture the transferable RS knowledge.

\section*{Acknowledgments}
This work was partially supported by National Key R\&D Program of China (No.2022ZD0118201), National Natural Science Foundation of China (No. 62002303, 42171456) and  Natural Science Foundation of Fujian Province of China (No. 2020J05001).

\bibliography{main}

\end{document}